# Operation and performance of a pilot HELYCON cosmic ray telescope with 3 stations

T. Avgitas[ab], G. Bourlis[a], G.K. Fanourakis[c], I. Gkialas[d], A. Leisos[a†], I. Manthos[de], A. Stamelakis[a], A.G. Tsirigotis[a] and S.E. Tzamarias[ae]

*a Physics Laboratory, School of Science & Technology, Hellenic Open University Patras, Greece*

*b Laboratory APC, University Paris Diderot - Paris VII, Paris, France*

*c Institute of Nuclear and Particle Physics, NCSR Demokritos, Athens, Greece*

*d Department of Financial and Management Engineering, University of the Aegean, Chios, Greece*

*e Department of Physics, Aristotle University of Thessaloniki, Thessaloniki, Greece*

## Abstract

Three autonomous HELYCON stations have been installed, calibrated and operated at the Hellenic Open University campus, detecting cosmic ray air showers. A software package for the detailed simulation of the detectors' response and the stations' operation has been developed. In this work we present the results of the analysis of the data collected by the stations during a period of one year and a half. The performance of the telescope is compared and found in very good agreement with the predictions of the simulation package. The angular resolution of each autonomous station is 3 to 5 degrees depending on the station geometry. In addition, by analyzing data from showers detected synchronously by more than one station, we evaluate the performance of the telescope in detecting very high energy ($E > 5 \cdot 10^{15}$ eV) cosmic rays.

## Introduction

We have previously presented [1] the construction, calibration, testing and evaluation of the performance of large charged particle detectors (plastic scintillators). We have also presented [2] the installation of three autonomous stations of such detectors, for the detection of Extensive Air Showers (EAS), at the campus of the Hellenic Open University (HOU), forming a pilot phase of the HEllenic LYceum Cosmic Observatories Network (HELYCON) [3]. The installation of the stations and their calibration procedures, as well as the developed Monte Carlo (MC) simulation package, the analysis of the calibration data and the comparison with the MC predictions and the fine tuning of the simulation package parameters have also been presented [2].

In this work, we present results from the analysis of the data collected during an operating period of more than a year and a half (from 2014 August 1st, to 2016 March 16th). These data sum up to a total of more than 600000 events collected by the three deployed autonomous stations and the analysis results are presented in comparison with the predictions of a detailed MC simulation. In section 1, we briefly outline the setup and the operation of the detectors network array. In section 2, we describe the developed MC simulation package, the data selection criteria and the procedures for the analysis of the events (experimental and MC). In section 3, we present results of the data analysis, concerning the detector signal processing and the reconstruction of the EAS direction from single autonomous stations. Finally, in section 4 we present results from the analysis of data that correspond to EAS recorded synchronously by more than one station.

---

† Corresponding author

## 1 Installation, calibration and operation of the detector array

The HELYCON array comprises three autonomous detection stations, located as depicted in Figure 1, each one consisting of three plastic scintillator detectors and a CODALEMA type RF antenna [4]. The typical distance of the scintillator detectors in a station is almost 27 meters forming a triangle, with the RF antenna in the middle of each station. An approximate equilateral triangle is formed in stations A and C, while the site constraints dictated station B to form an amblygonal triangle, offering the opportunity to study the efficiency and geometrical acceptance of such a geometry.

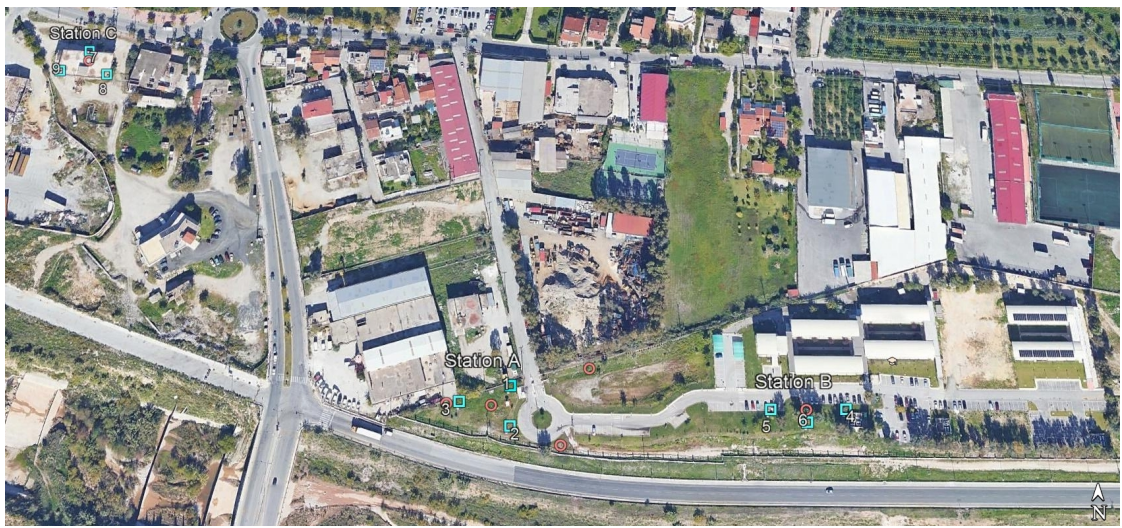

Figure 1: The HELYCON array installed at the Hellenic Open University campus. The geodesic coordinates of the stations centers are: Station A (38°12'22.69"N, 21°45'52.60"E), Station B (38°12'22.49"N, 21°45'59.32"E), Station C (38°12'29.62"N, 21°45'42.36"E).

As described in [2], a special, metallic, waterproof box hosts the essential equipment for powering the detectors and the antenna, the slow control system and the data acquisition system, which is based on the Quarknet DAQ board [5]. The Quarknet board registers the times of the crossings of the waveforms with an adjustable threshold, allowing the timing of the pulses and the estimation of their size using the Time over Threshold value (ToT). A local computer is used for the control and monitor of the station and the data acquisition, as well as for the communication with the Physics laboratory main server. The calibration of the photomultipliers and the detectors [1], as well as the calibration of the stations, was performed employing several experimental setups and the response of the detectors (including the effect of the signal transmission cables) is perfectly understood and described in the simulation [2]. In addition, the parameterization of the pulse characteristics, such as charge and peak voltage, as a function of ToT values has been evaluated by means of the MC simulation [2].

The DAQ system of each station is triggered by the 3-fold coincidence of the particle detectors' signals. A preselected threshold of 9.7 mV (around two times the pulse height of a MIP (Minimum Ionizing Particle)) is employed, common for all detectors and the time window for the coincidence is adjusted to take into account the distances between the detectors (typically 150 ns). Each time such a coincidence occurs, the DAQ system produces a NIM pulse (Quarknet-OUT) that triggers the RF antenna of the station. This NIM pulse is produced with an equal probable jitter of 10 ns. Both the Quarknet board and the antenna are equipped with a GPS receiver, for the time tagging of each recorded event, allowing for offline identification of event coincidences among different stations. The analysis of the antenna data was performed independently and it is presented separately [6].

## 2 Simulation framework and event selection

### 2.1 Simulation framework

The simulation of the HELYCON detectors is a two-step process. At first, the CORSIKA simulation package [7] simulates the evolution of the EAS, down to the detector level. The QGSJET-II-04 [8] and GEISHA [9] packages have been employed for high and low energy hadronic interactions respectively and the EGS4 Code system [10] for electromagnetic interactions. For the primary particles energy, direction and composition, the standard distribution was used [11] and the atmosphere was modeled using the Atmospheric model 7, which corresponds to the central European atmosphere for Oct. 14, 1993. We have produced $2\cdot10^8$ EAS with primary particle energy in the range $10^{13}-5\cdot10^{15}$ eV and $10^5$ EAS between $5\cdot10^{15}$ and $10^{18}$ eV, registering for each primary particle the energy and the arrival direction, as well as, for each particle of the shower, the position and the arrival time at the detector level. In order to increase the statistics of the MC sample, each CORSIKA shower was used more than once, by moving the impact point uniformly inside a large enough circular area around the center of each station[1]. The MC sample equivalent lifetime was 15300 h for each station and 112000 h for multiple stations.

In the second simulation step, the Hellenic Open University Reconstruction and Simulation (HOURS) package [12] was employed to simulate the response of the detectors to EAS particles (scintillating material, optical fibers, photomultipliers, cable effects). In addition, the functionality of the DAQ system was simulated in detail and the times of the crossings of the waveforms with the threshold were calculated and stored in the same format as that of the raw data. The following analysis steps were applied to both the experimental data and the simulated events.

### 2.2 Event Selection

The identification and rejection of events caused by noise is of crucial importance,

---

[1] When studying single station performance the radius was chosen to be 420 m, since beyond this area only very high energy showers are detectable which, however, constitute a negligible fraction of the detected EAS. For multiple station events the corresponding radius was 670 m and only high energy events were used since low energy EAS particles cannot trigger more than one station.

especially in station B, which is deployed next to a strong man-made noise source. Consequently, quality criteria were applied to all detected signals in order to be validated as events of cosmic origin. As depicted in Figure 2 the man-made noise in the telescope area appears in clusters of very small time intervals between successive events. Thus, events with inter-event time less than 0.1 s in each station were rejected. Other selection criteria included the demand for alternating rising and falling edges[2], at least one rising and one falling edge on each waveform, while unexpected edges (e.g. a rising edge following a rising edge) were disregarded and a recheck of the above criteria was performed.

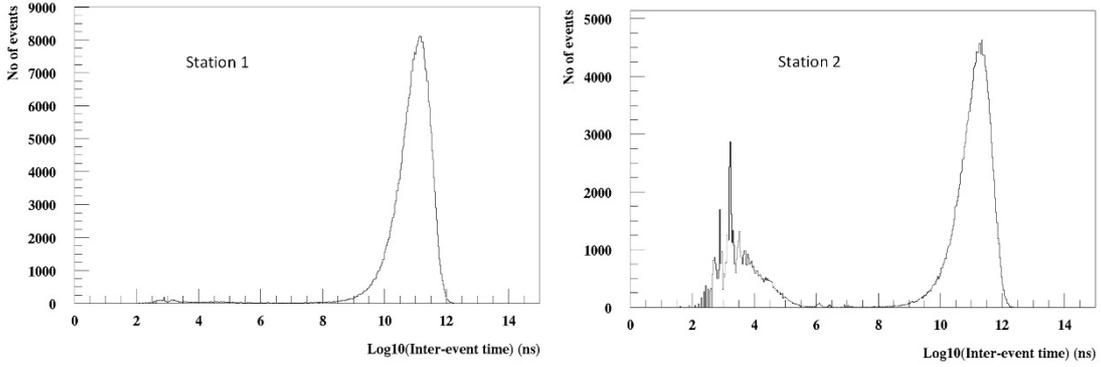

Figure 2: Distribution of time difference, in logarithmic time scale, between consecutive events. The different behavior on station B is due to the noise clusters.

For each detector waveform, the surviving pulses were merged if all the following conditions were met: a) merging was performed only within a time window of 500 ns in order to avoid the effect of cable reflections, b) only following smaller pulses were incorporated to bigger pulses and c) the time difference between successive pulses was smaller than 30 ns. If after merging there remained more than one pulse, only the first pulse was used in the subsequent steps of the analysis. In order to improve data quality, small width pulses (ToT value less than 15 ns) were rejected. In Table 1 the numbers of the selected events, the quality passed events, as well as the operating period of each station is listed. It should be noted that the smaller number of events in station 2 is expected due to the geometry of the station and the shorter operation time.

| Station | A | B | C |
| --- | --- | --- | --- |
| No of selected events | 338658 | 245275 | 368804 |
| No of quality passed events | 226269 | 116850 | 249570 |
| Operating period (h) | 12946 | 10123 | 13222 |

Table 1: Number of experimentally selected events in each station before and after the application of noise rejection. The effect of the selection criteria in station B, clearly shows the increased noise level already mentioned and depicted in Figure 2. Additionally, the operating period of each station is presented.

After selecting events fulfilling these quality criteria, a pulse timing correction as a function of ToT was utilized, as described and parameterized in [2]. In addition, taking into account the results of the calibration of the stations (parameterizations of the pulse characteristics versus the ToT values), the pulse characteristics as well as the EAS local shower front direction (azimuth and zenith angle) were estimated. In addition, for MC events the collected charge in each detector (calculated from the integral of the full waveform), the direction and the primary energy of the EAS as well as the impact point of the shower core at the detectors level were saved.

---
[2] Crossings of the pulses with the predefined threshold level.

## 3 Single station results

### 3.1 Data analysis and MC comparison

Although the only directly available experimental information is the times of the crossings of the waveforms with the threshold (allowing the timing of the pulses and the estimation of their size using the ToT values), more distributions using the parametrizations versus the ToT value can be compared with the MC predictions. The ToT distribution for each detector of station A is demonstrated in Figure 3 (left) for data and MC, while the distribution of the sum of the ToT values of the three detectors of the station is depicted in Figure 3 (right) and it is found to be in good agreement with the corresponding MC estimation. The depicted ToT distributions of the MC events have been normalized to the operating period (presented in Table 1 for each station).

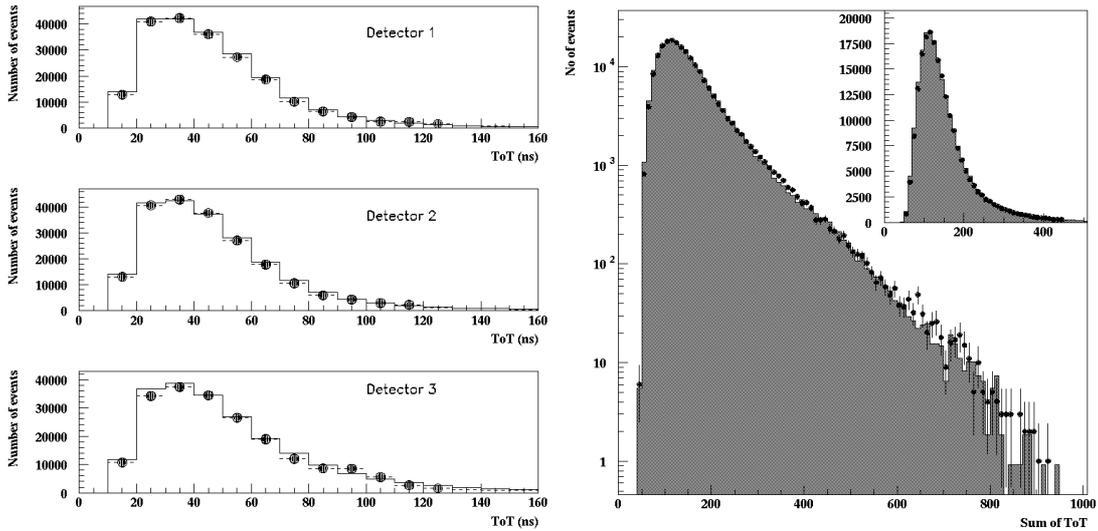

Figure 3: Left: Distribution of the ToT values for each detector of station A, for experimental data (dots) and MC estimation (histogram), normalized to the operation time of the detecting station. (Right) The sum of the three ToT values for each event in semi-log scale, for experimental data (dots) in comparison with MC simulation (histogram) normalized to the operation time of the detecting station. Same plot in linear scale appears in the inset plot.

Using the charge versus ToT parameterization [3], we can estimate the pulse charge (in units of the MIP equivalent charge) for experimental data and MC events. For the MC events, the charge is additionally calculated from the full waveform, which is available in the simulation, allowing thus a more accurate comparison. The charge distributions for station A are depicted in Figure 4, in the same way as for the ToT value distributions of the previous Figure.

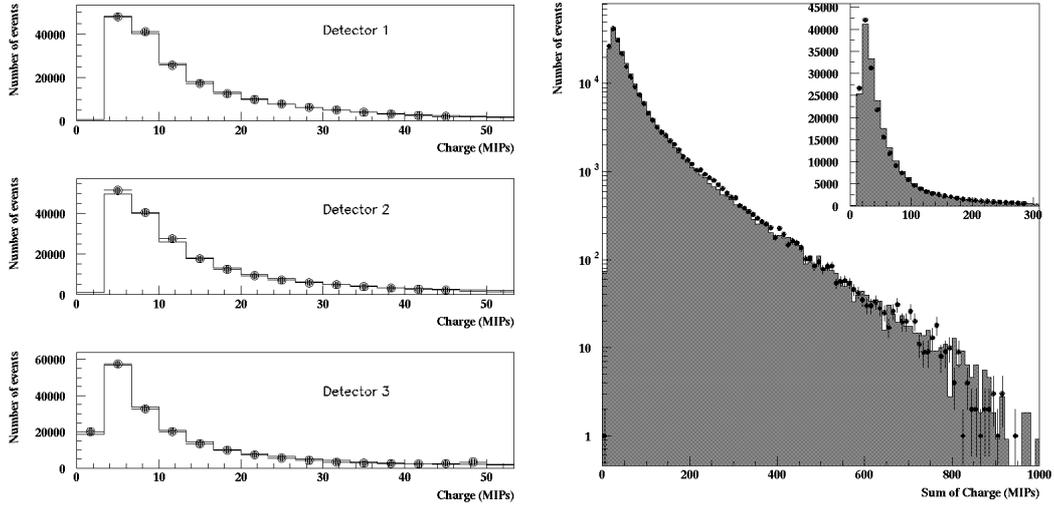

Figure 4: Distribution of the charge of each detector of station A, as estimated from the charge versus ToT parameterization for experimental data (dots) and as calculated from the full waveform of the MC estimation (histogram), normalized to the operation time of the detecting station (left). Distribution of the sum of the charge of the three detectors for each event in semi-log scale (right), for experimental data (dots) in comparison with MC prediction (histogram). Same plot in linear scale appears in the inset.

The above ToT and charge distributions of experimental data and MC events have been generated for all the HELYCON stations [13]. The mean values and rms of these distributions for all stations appear in Table 2.

| Station | Detector No | ToT | | Charge | |
|---|---|---|---|---|---|
| | | Mean value (Data – MC) (ns) | rms (Data – MC) (ns) | Mean value (Data – MC) (MIP equiv.) | rms (Data – MC) (MIP equiv.) |
| A | 1 | 52.91 – 52.16 | 35.27 – 33.71 | 29.40 – 27.11 | 47.81 – 42.95 |
| | 2 | 52.26 – 52.58 | 34.61 – 34.32 | 27.21 – 27.27 | 44.67 – 43.32 |
| | 3 | 64.93 – 64.93 | 41.77 – 42.49 | 27.25 – 26.17 | 39.22 – 37.31 |
| B | 1 | 46.70 – 48.49 | 33.09 – 34.55 | 27.24 – 25.93 | 47.99 – 45.89 |
| | 2 | 47.70 – 50.24 | 36.43 – 38.23 | 23.85 – 24.26 | 42.57 – 42.46 |
| | 3 | 57.81 – 57.29 | 37.48 – 35.75 | 33.69 – 31.56 | 50.87 – 45.28 |
| C | 1 | 48.42 – 48.50 | 32.53 – 32.11 | 25.67 – 24.96 | 46.74 – 40.60 |
| | 2 | 51.19 – 53.08 | 36.71 – 38.70 | 22.83 – 22.58 | 39.77 – 37.61 |
| | 3 | 47.60 – 48.82 | 33.92 – 34.56 | 24.04 – 23.60 | 44.80 – 40.49 |

Table 2: Mean and rms of the ToT and charge distributions for each detector of the 3 HELYCON stations for both experimental data and MC estimation.

Reconstruction of the EAS local shower front direction (zenith and azimuth angles) is realized with the triangulation method. The arrival time of the pulse, that is the time of the first crossing of the pulse waveform with the threshold, corrected for timing systematical errors (slewing), is used for this reconstruction, taking also into account possible time differences due to the different lengths of the signal cables.

The distributions of the reconstructed angles are presented in Figure 5 for station A, in comparison with the distributions of the reconstructed angles of the MC events.

It should be noted that the reconstructed direction of the local shower front[3] deviates from the direction of the primary particle initiating the shower simulated by CORSIKA.

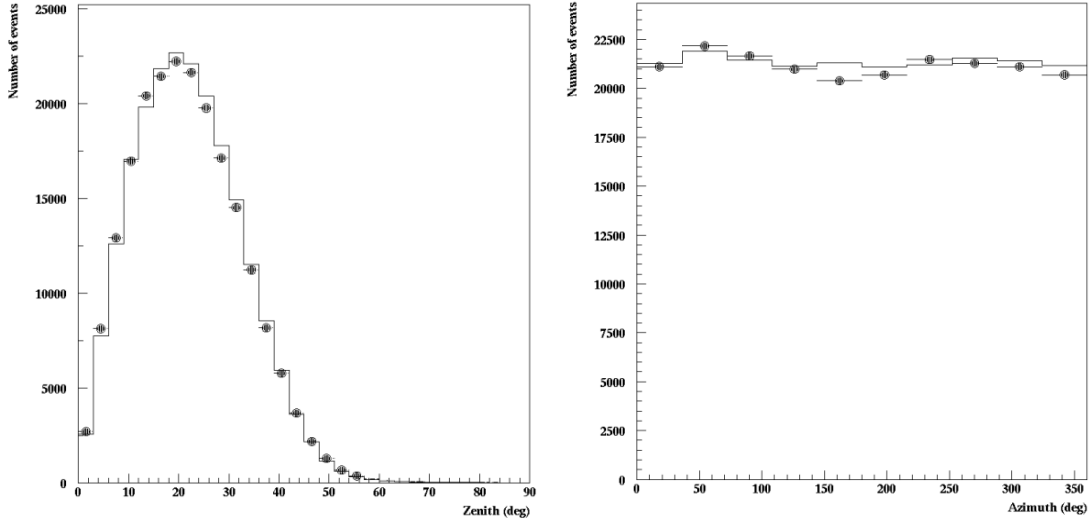

Figure 5: Distributions of the zenith (left) and azimuth (right) angles of the reconstructed direction of EAS local shower front. The experimental data are represented with dots, while the MC prediction with the histogram.

Table 3 contains information (mean value and rms) for the distributions of Figure 5 for station A and additionally for the other two HELYCON stations [13]. The azimuth angle follows an equal-probable distribution as expected, while the zenith angle distribution is fitted with a model flux of the form

$$\frac{dN}{dcos\theta} \sim cos\theta^{\alpha} \qquad (1)$$

The fit gives a spectral index of $\alpha = 9.55 \pm 0.02$ in agreement with other similar measurements [14] and is depicted in Figure 6.

| Station | Zenith angle (data – simulation) (in degrees) | | Azimuth angle (in degrees) |
|---|---|---|---|
| | Mean value | rms | Mean value |
| A | 21.85 – 21.95 | 10.69 – 10.50 | 179.0 – 179.7 |
| B | 22.24 – 22.28 | 11.24 – 11.67 | 180.5 – 185.9 |
| C | 21.81 – 21.66 | 10.77 – 10.48 | 177.2 – 178.8 |

Table 3: Mean value and rms for the zenith angle distribution and mean value of the azimuth angle distribution, as reconstructed from the experimental data in comparison with the reconstructed angle values of the simulation's estimations, for each HELYCON station.

The zenith and azimuth angle distributions for Station B are slightly different compared to the respective distributions of the other stations, due to the station geometry. However, the agreement between experimental data and the simulation estimation is almost perfect, as can be seen from Figures 4 and 5 (for station A), as well as from Tables 2 and 3 (for all stations).

---

[3] Due to the curvature of the shower front and the relative small size of the station, the reconstructed direction using triangulation corresponds to the direction of the part of the shower front incident on the station (local shower front).

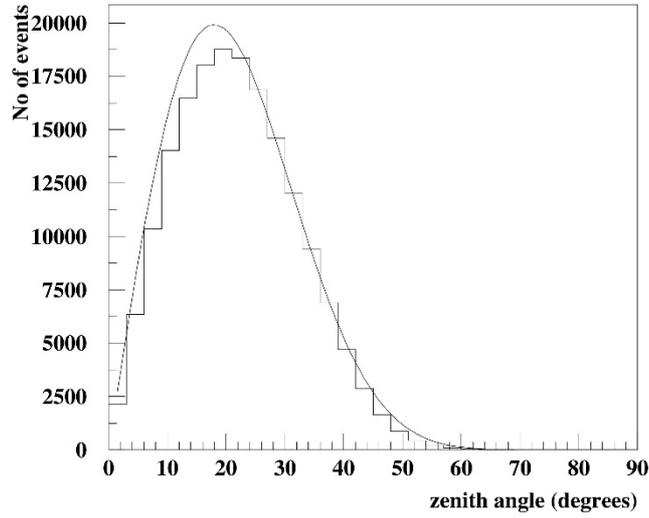

Figure 6: Distribution of the reconstructed zenith angle of EAS local shower front, fitted with a model flux of the form of eq. (1).

### 3.2 Simulation study

The observed agreement between the experimental data distributions and the corresponding MC distributions allows us to study the station performance using the full MC information (like the impact point of the EAS and the energy of the primary particle). The full simulation study has been performed for all three HELYCON stations [13], but in this paper results concerning station A will be presented with the exception of some results explaining station B behavior.

Using MC events we estimate the error in the direction (zenith and azimuth angle) of the primary particle as the sigma of the Gaussian distribution of the difference between the angle of the EAS primary particle and the reconstructed angle of the EAS local shower front. In Figure 7, the resolution of the EAS shower front direction estimation is presented and in particular the error of the zenith angle estimation (left), the error of the azimuth angle estimation (middle) and the median angle between true and reconstructed direction of the shower (right), in bins of the distance of the EAS impact point from station A center (top), in bins of the sum of the ToT values of the three detectors for station A (center) and versus the total charge collected by all three detectors of station A (bottom). For all angles an increase of the estimated error with the increase of the distance is observed. This is explained taking into account the fact that the EAS shower front is approximated with a plane perpendicular to the shower axis, ignoring the curvature, and this has an increasing effect as the EAS impact point moves away from the station, introducing time differences not taken into account in the reconstruction. Additionally, EAS incident at greater distances from the station, tend to cause smaller detector pulses, suffering from larger timing errors, with the respective influence on the resolution of the reconstruction. In the case of evaluating the reconstructed angles resolution in bins of ToT, there is a reduction of the error as the measured ToT values increase. This is justified as higher pulses are demonstrating smaller timing errors, in addition to the fact that higher pulses are indicating EAS with impact points closer to the station, in agreement with what is mentioned above.

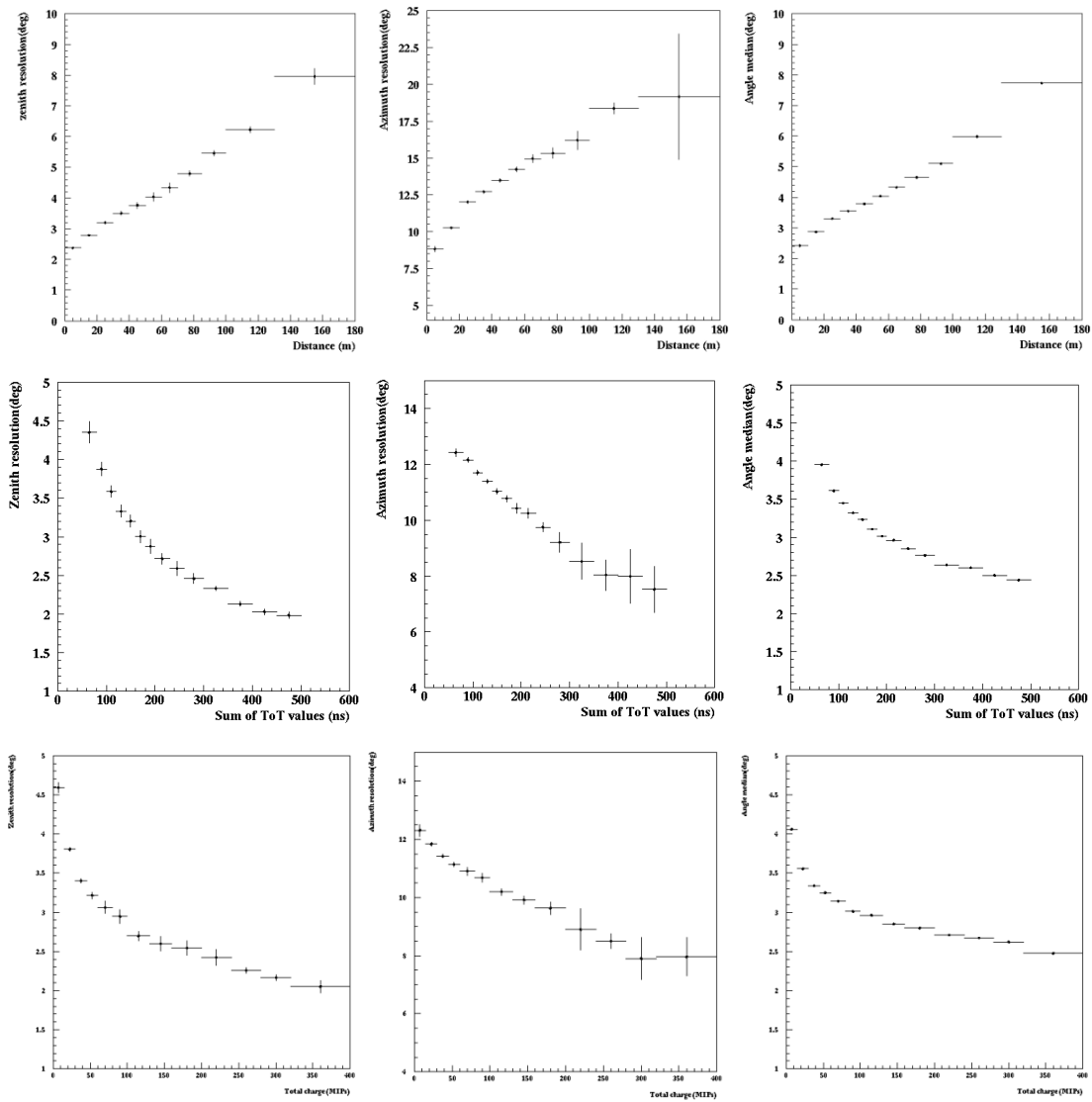

Figure 7: Error of the zenith (left) and the azimuth angle estimation (middle) and median angle between true and reconstructed direction of the shower (right), in bins of the distance of the EAS impact point from the center of station A (top), in bins of the total ToT (sum of the ToT of the three detectors) (center) and versus the sum of the charge of the three detectors of station A (bottom).

In Figure 8 the estimated error of the reconstructed zenith angle (top) and azimuth angle (bottom) is presented as a function of the zenith (left) and the azimuth angle (right). In the case of the zenith angle resolution as a function of the zenith angle, an increase of the estimated error is noted as the zenith angle grows. This is due to the reduction of the effective area of a flat detector as a function of $cos\theta$. Moreover, EAS of larger zenith angles go through greater distances inside the atmosphere, resulting in fewer particles reaching the detectors. This is more obvious if the events of the same bin are grouped according to their charge, when a better resolution is observed for EAS of higher charge (Figure 8 top left). Considering the error estimation of the azimuth angle as a function of the zenith angle, a decrease of the error is observed with increasing zenith angle. This can be justified as particles of EAS with greater zenith angle values impinge to the station detectors with greater time differences, resulting in a more accurate azimuth angle reconstruction (an effect with greater impact than the reduction of charge due to the inclination of the shower).

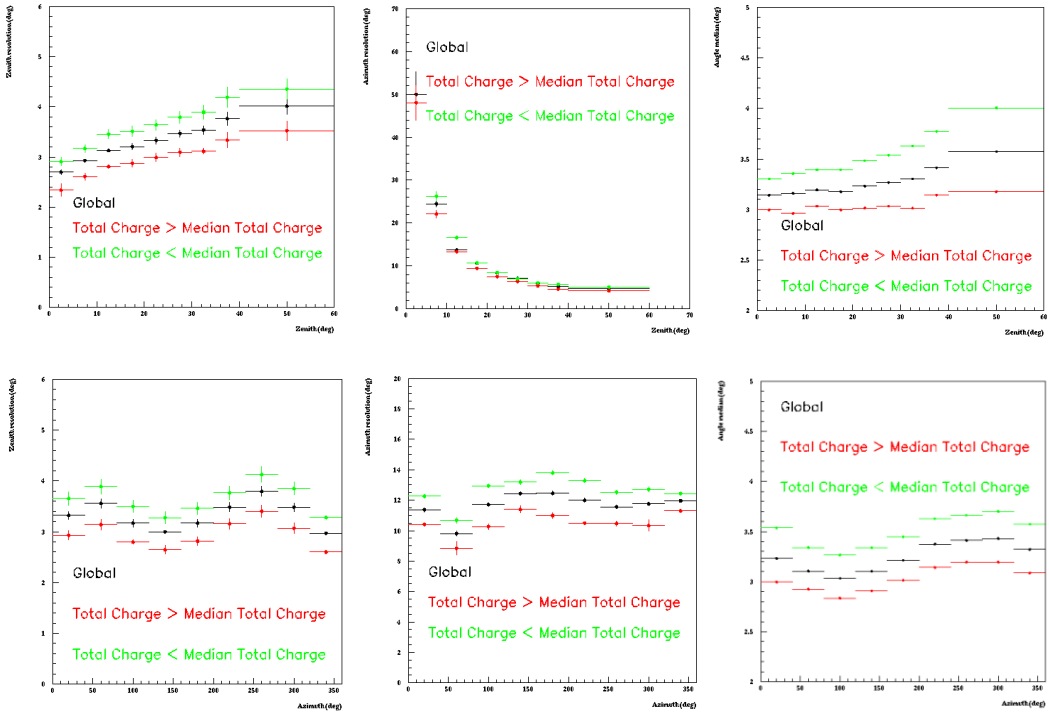

Figure 8: Error of the zenith angle (left) and of the azimuth angle estimation (middle) and median angle between true and reconstructed direction of the shower (right), versus the zenith angle (top) and the azimuth angle of the EAS (bottom) for station A. The plots in green and red correspond to a collected charge above and below the median of the total charge distribution respectively, while all the reconstructed events are included in black.

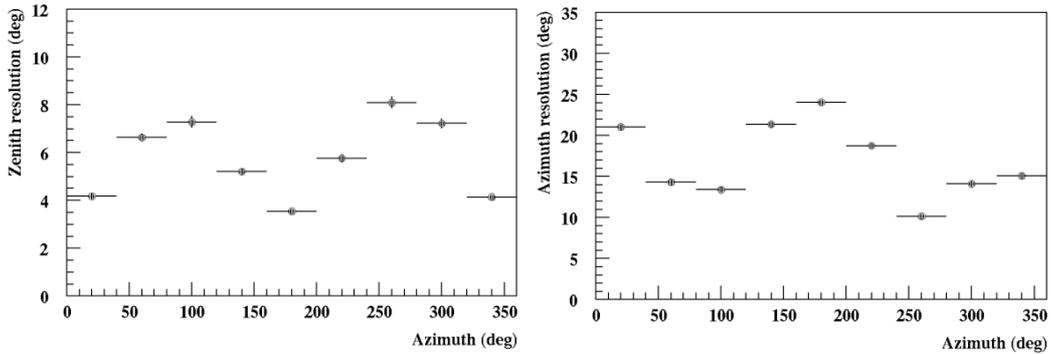

Figure 9: Error in zenith (left) and azimuth (right) angle estimation in bins of the azimuth angle of the EAS for station B.

Concerning the error estimation of the zenith and azimuth angles as a function of the azimuth angle, a deviation from a flat distribution is observed, which, as can be seen in Figure 9, is stronger for station B and is due to the geometry of the stations. By implementing a "toy" Monte Carlo, including only the influence of the detectors position, the same behavior is noted for both stations, attributed to the orientation and positioning of the particle detectors within each station, as depicted in Figure 10 for station A (top) and station B (bottom). The peak (valley) for the zenith (azimuth) resolution around 90 and 270 degrees is justified by the lever arm of the station which is elongated in the east-west direction for station B. This effect is also present for station A, but to a lesser degree.

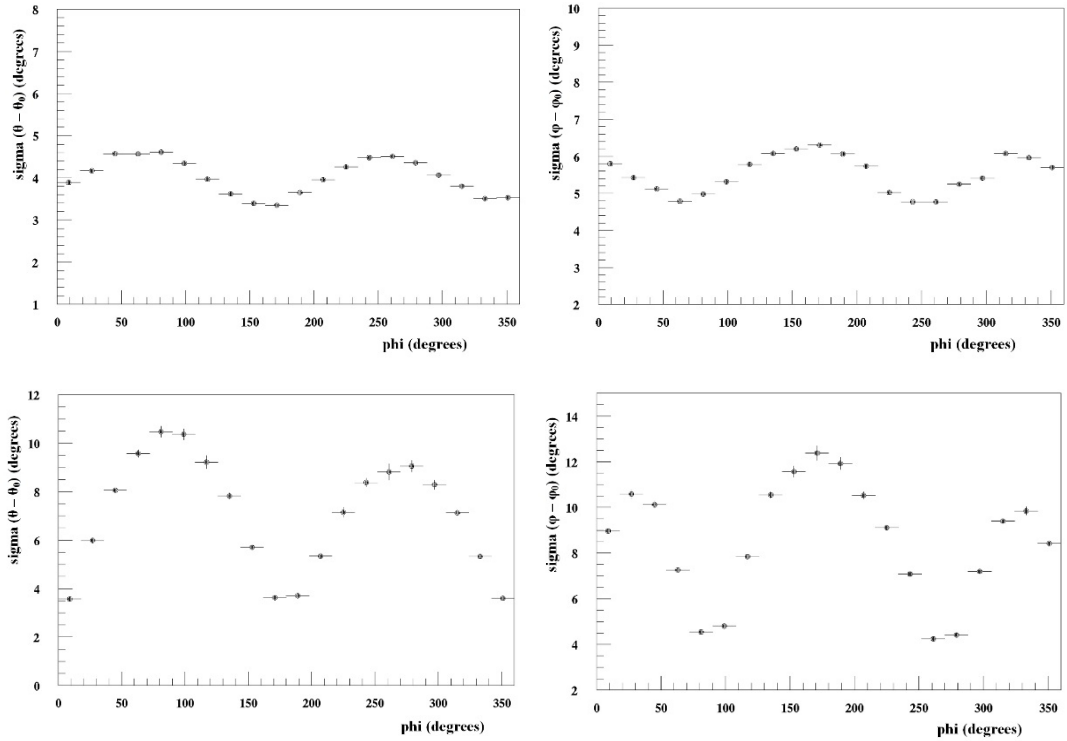

Figure 10: Error in zenith (left) and azimuth (right) angle estimations in bins of the azimuth angle of the EAS for station A (top) and station B (bottom) detectors, as resulted from the "toy" MC simulation.

Figure 11 depicts the dependence of the zenith angle reconstruction error on the energy of the EAS primary particle for all EAS (black). This dependence is also plotted for EAS with an impact point that lays more (red) or less (green) than 20 m from the station center. Reconstruction error for EAS with impact points close to the station is, as expected, decreasing with increasing energy. On the other hand, there is an increasing number of detected EAS with their impact point in large distances from the center of the station (in red) as the energy increases. For these EAS the effect of the shower front curvature is larger, as already mentioned, and they possibly have lower ToT values, thus larger timing errors. Since these EAS are dominating the total number of events (in black) an increasing zenith angle estimation error is observed.

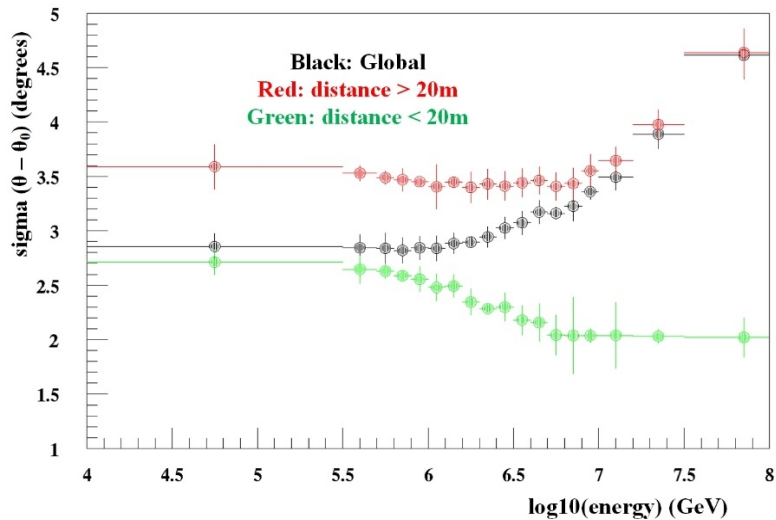

Figure 11: Error in the zenith angle reconstruction estimation in bins of the EAS primary particle

energy for station A. Black points are for all MC reconstructed events, while green and red points correspond to showers with impact points less or more than 20 m from the station center respectively.

Figure 12 shows the angular difference between the true primary particle direction, and the reconstructed direction employing the triangulation method and the information of the pulses timing for station A. The difference between the reconstructed EAS local shower front direction and the true direction of the EAS has a median value of 3.25 degrees for station A. The corresponding median value for station B is 5.3 degrees, due to the station geometry while for station C it is 3.5 degrees.

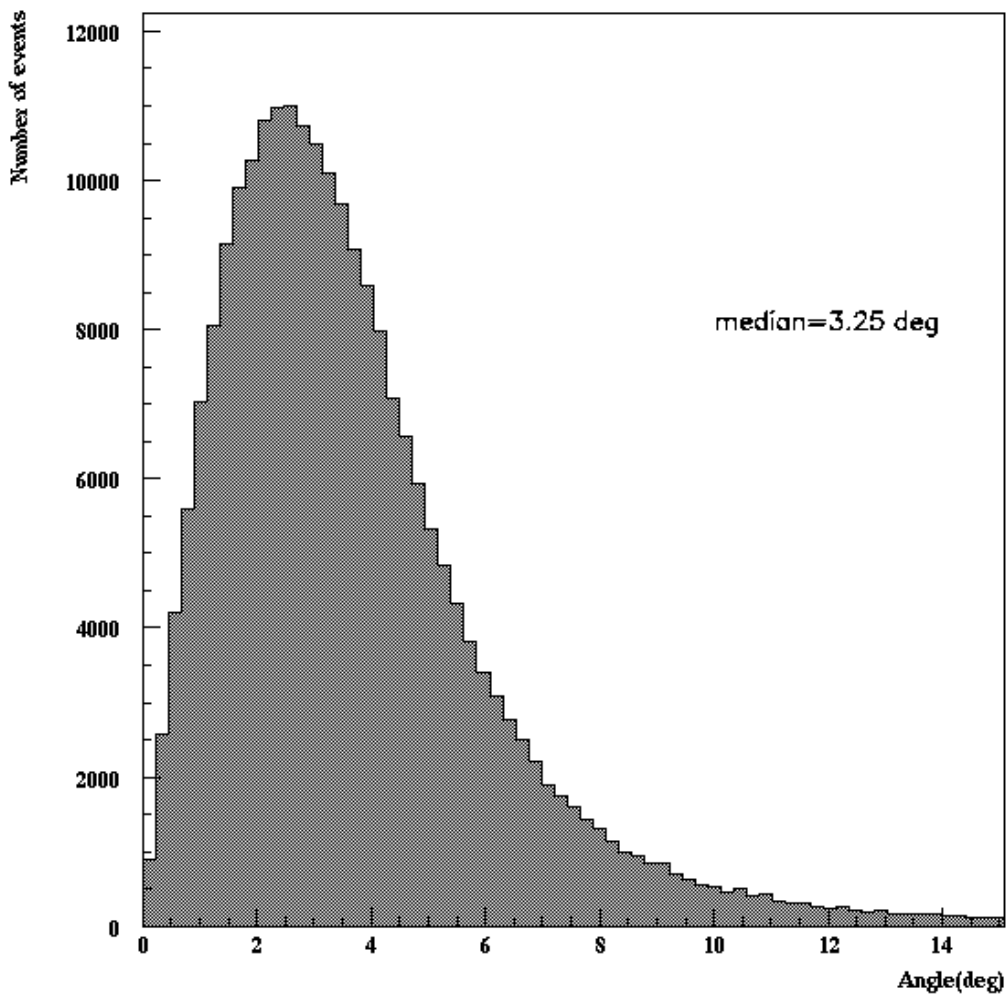

Figure 12: Angular difference between true EAS direction and EAS local shower front direction as reconstructed by the triangulation method, using the timing of the pulses of station A.

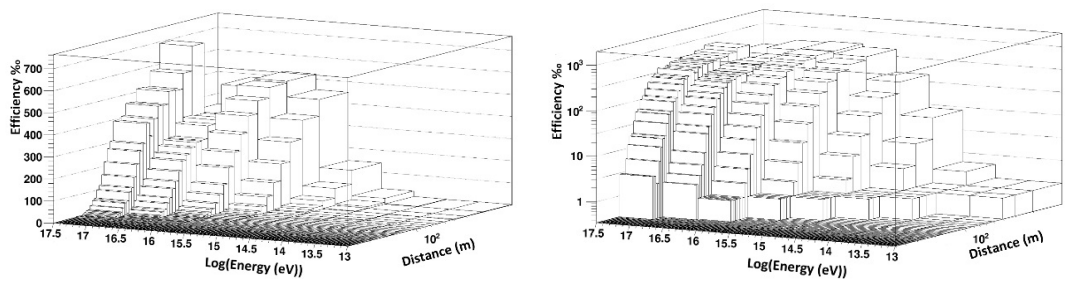

Figure 13: Efficiency of station A as a function of the EAS impact point distance from the center of the station and the EAS primary particle energy in linear scale (left) and logarithmic scale (right).

The station efficiency in detecting EAS depends on the energy of the primary particle and on the distance of the EAS impact point from the station center. The efficiency of station A versus the distance of the impact point from the station center and the primary energy of the EAS, is presented in Figure 13 in linear scale (left) and logarithmic scale (right) for the efficiency axis.

The efficiency of station A for events close to the station and for the highest energies simulated, exceeds 60% and falls to zero at the borders of the simulation area. The maximum efficiency for station B is about 30% due to the geometry of the station making the formation of the desired 3-fold coincidence level harder, while the maximum efficiency for station C is approximately 60%, close to that of station A.

## 4 Multiple station coincidence results

The distance between stations A and B (164 m) is the smallest between the HELYCON stations and consequently this combination offers a satisfactory number of double station coincidence events for analysis. In particular 1395 experimental events in an operation period of 9402 h were selected and successfully reconstructed. This number is consistent with the 16803 MC events selected and successfully reconstructed in an equivalent operation time of approximately 112000 h (corresponding to 1410 events when normalized to the experimental operation time). Table 4 presents the numbers of detected and expected events for every double and triple station combination.

| Stations | Detected events | Expected events | Operation time (h) |
|----------|-----------------|-----------------|--------------------|
| A-B      | 1395            | 1410            | 9402               |
| A-C      | 33              | 30              | 12288              |
| B-C      | 12              | 12              | 9450               |
| A-B-C    | 6               | 8               | 8904               |

Table 4: Number of experimentally detected events in comparison with the expected from the simulation for each double and triple station combination.

The selection of these events is realized by checking whether the absolute GPS times of all stations fall within a window of 1500 ns, which is more than enough even for horizontal showers, taking into account the distance between the stations.

The number of multiple station coincidence events is much smaller than the number of single station events since only large higher energy showers can trigger more than one station. This is demonstrated in Figure 14 where the energy distribution of the EAS primary particle is depicted for showers detected by station A and for showers detected by both stations A and B.

Due to the large distance between stations A and B, the impact points of the reconstructed showers are distributed mainly in the region between the two stations (Figure 15, left). The detector of station B that is positioned closer to the middle of the distance between stations A and B (that is Detector 5), is expected to fire earlier for most of the recorded events. Figure 15, right depicts the number of events for which each of the 6 detectors of stations A and B exhibits the earliest signal. In agreement with the MC expectations, Detector 5 exhibits the earliest signal more frequently.

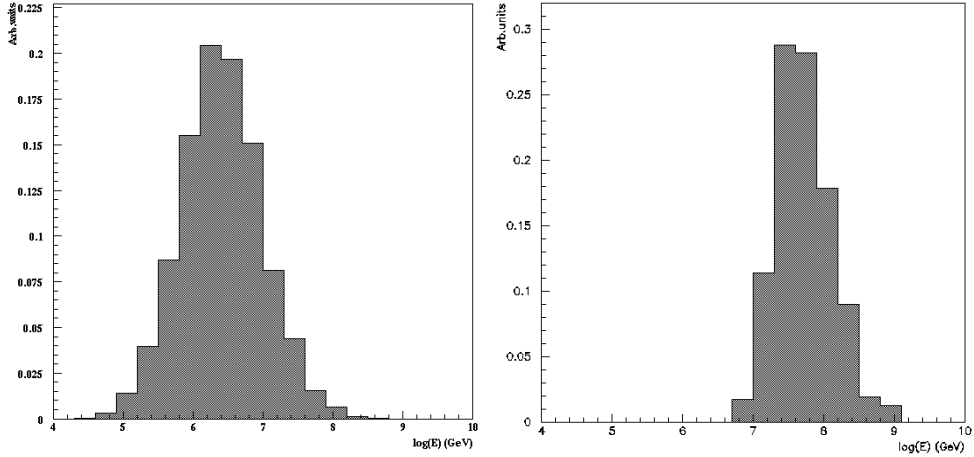

Figure 14: Primary particle energy distribution of EAS reconstructed from station A (left) and stations A and B in coincidence (right).

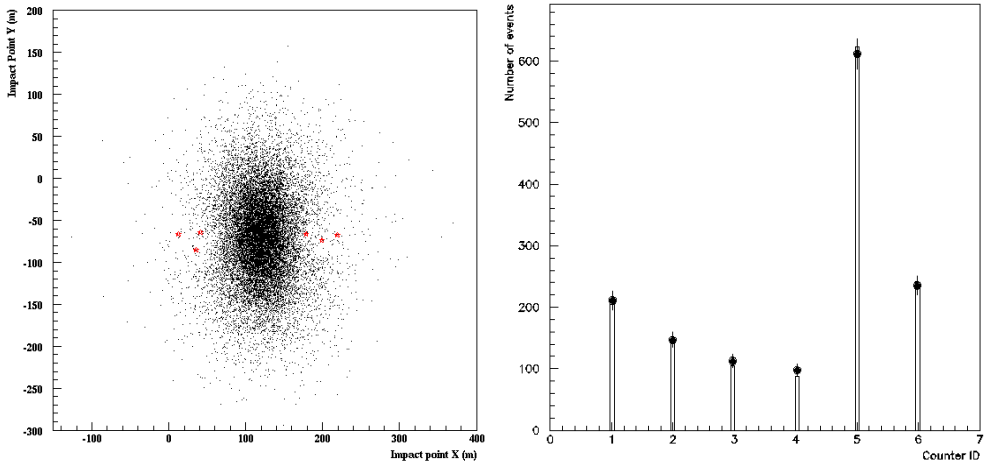

Figure 15: Spatial distribution of the impact points of the reconstructed showers, detected by both stations A and B (left). Red stars indicate the positions of the stations detectors. Number of events with the respective detector exhibiting the earliest signal among the 6 detectors of stations A and B (right). Dots correspond to experimental data and bars to the simulation results.

As mentioned the energy threshold of the primary particle is shifted to higher values compared to the single station recorded showers. Experimentally, this is demonstrated in the distribution of the mean of the ToT values of the recorded waveforms in a station, which is shifted to higher values compared to the corresponding distribution of single station recorded showers (Figure 16).

Even though for the double station coincidence showers the deposited charge is larger, the performance of each station separately is getting worse with respect to single station operation (section 3). For example in station A the zenith angle resolution is on average 5.60 degrees compared to 3 degrees on average (see Figure 8 (top – left) for zenith angle 22 degrees, which is the mean reconstructed zenith angle as depicted in Table 3). This is due to the large distance between station A and the impact point of double coincidence showers. For the same reason reconstructed showers from stations A and B in coincidence are more inclined than the corresponding showers recorded from a single station.

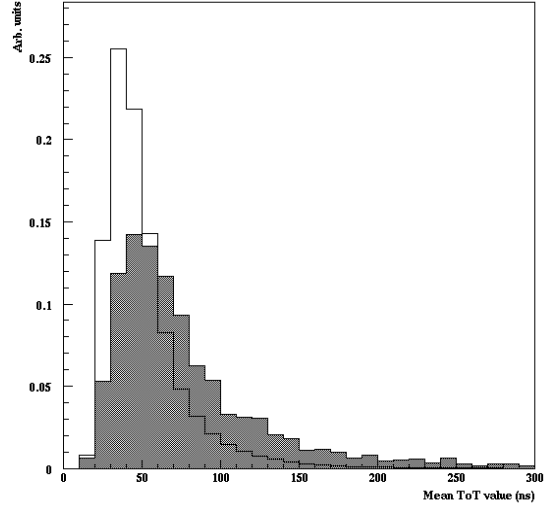

Figure 16: Distribution of the mean ToT value of the detectors for reconstructed events detected only by station A (plain histogram) and for those detected by both stations A and B (shaded histogram).

The reconstruction of the angular parameters of the showers that trigger both stations (that means a 6-fold coincidence of the scintillator detectors) has to take into account the curvature of the shower front. Even though in single station operation the effect is small due to the small inter-detector distance, the difference of the arrival time of the shower front between two stations deviates from the predicted value of the plane particle front approximation. This deviation depends on the distance of the stations from the shower axis. If one station is very close to the shower axis and the other far away the effect is big (15-20 ns), while when both stations are nearly at the same distance from the shower axis, the effect is negligible (due to the symmetry of the shower front with respect to the shower axis). If we consider the most energetic pair of counters ($A_{max}$, $B_{max}$) from the two stations A and B (where $A_{max}$ and $B_{max}$ are the counters with the highest collected charge among the counters of stations A and B respectively) and calculate the distances $dA_{max}$ and $dB_{max}$ of these counters from the shower axis, then the ratio

$$f = 100 \cdot min(dA_{max}, dB_{max}) / max(dA_{max}, dB_{max}) \qquad (2)$$

can be used to select showers that are either close to one station or in between. For example in Figure 17 the positions of the impact points of the showers used in the simulation are shown for 3 different values of f: green (f<20), blue (30<f<50) and black (f>90). The positions of the counters are depicted with red color.

In order to estimate the ratio f, we use the ratio

$$fq = 100 \cdot min(qA_{max}, qB_{max}) / max(qA_{max}, qB_{max}) \qquad (3)$$

where $min(qA_{max}, qB_{max})$ and $max(qA_{max}, qB_{max})$ is the smallest and largest charge (normalized in MIPs) respectively, collected by the most energetic detector of stations

A and B, per unit area perpendicular to the shower axis. In Figure 18 (left) the scatter plot of the ratio $f$ versus the ratio $fq$ is shown. Figure 18 (right) depicts the deviation $\Delta(fq) = dt_{true} - dt_{plane}$ as a function of the ratio $fq$, where $dt_{true} = t(max(qA_{max}, qB_{max})) - t(min(qA_{max}, qB_{max}))$ and $max(qA_{max}, qB_{max})$ and $min(qA_{max}, qB_{max})$ are defined as described above for equation (3) and $dt_{plane}$ is the expected time difference assuming plane particle front (depending only on the position of the two counters and the direction of the shower axis and not on the shower's impact point).

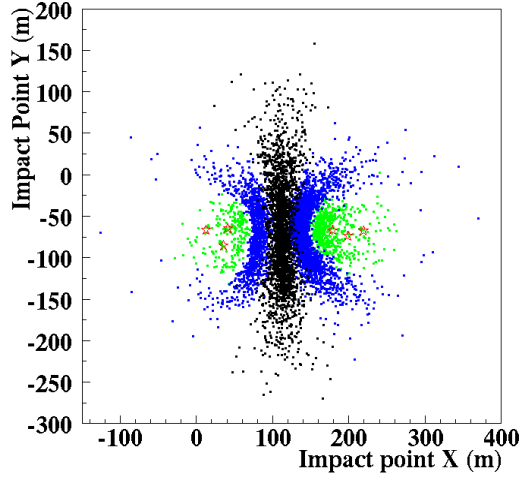

Figure 17: Positions of the impact points of the showers used in the simulation for 3 different values of the ratio f. Green is for $f < 20$, blue for $30 < f < 50$ and black for $f > 90$. Red stars indicate the positions of the detectors of stations A and B.

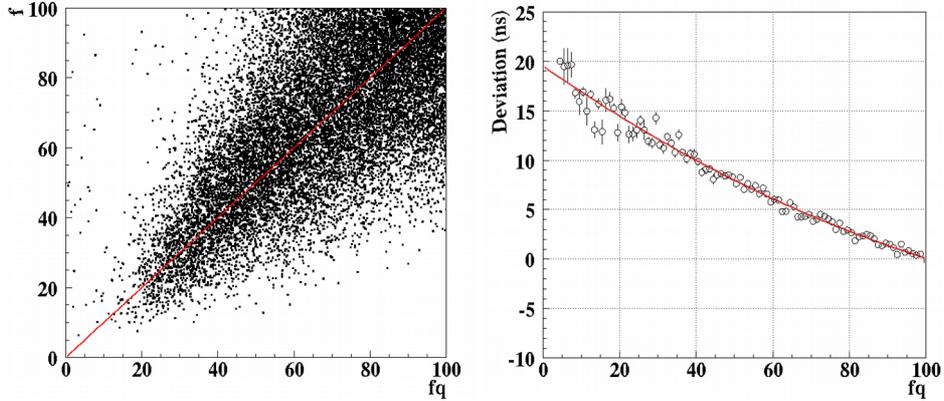

Figure 18: Scatter plot of the ratio f versus the ratio $fq$ (middle). Deviation $dt_{true} - dt_{plane}$ as a function of the ratio $fq$ (right).

The zenith and azimuth angles are reconstructed by determining the optimal combination of the directional angles (zenith and azimuth), which is achieved by minimizing, using the MINUIT package, the $\chi^2$ value of the function

$$\chi^2 = \sum_{iA=1}^{2} \left( \frac{dt_{iA, A_{max}}^{exp} - dt_{iA, A_{max}}^{meas}}{\sigma_{iA}} \right)^2 + \sum_{iB=1}^{2} \left( \frac{dt_{iB, B_{max}}^{exp} - dt_{iB, B_{max}}^{meas}}{\sigma_{iB}} \right)^2 + \\ + \left( \frac{dt_{A_{max}, B_{max}}^{exp} - \Delta(fq) - dt_{A_{max}, B_{max}}^{meas}}{\sigma_{AB}} \right)^2$$

(4)

where $dt_{iA,A_{max}}^{exp}$ and $dt_{iB,B_{max}}^{exp}$ are the expected time difference between the most energetic detector and the rest of the detectors for stations A and B respectively assuming a plane particle front, $dt_{iA,A_{max}}^{meas}$ and $dt_{iB,B_{max}}^{meas}$ are the corresponding measured time differences, $dt_{A_{max},B_{max}}^{exp}$ and $dt_{A_{max},B_{max}}^{meas}$ are the expected and measured time differences between the most energetic detectors of station A and B, while $\Delta(fq)$ is equal to the deviation described in the previous paragraph. $\sigma_{iA}$, $\sigma_{iB}$ and $\sigma_{AB}$ are the errors of the respective measurements including both the experimental resolution due to slewing as well as the statistical fluctuation of the particle front due to its thickness. $\sigma_{AB}$ includes the 10 ns equal-probable distribution jitter of the Quarknet trigger signal as well.

The distributions of the corresponding angles are presented in Figure 19, for the zenith angle on the left and for the azimuth angle on the right. Experimentally collected data are represented with dots, while the simulation estimation is represented by histograms. The mean value of the zenith angle distribution is 23.93 degrees for the experimental data and 24.42 degrees for the simulation estimation, with rms values of 11.22 degrees and 11.55 degrees respectively. For the azimuth angle distribution, the mean value is 178.4 degrees for the experimental data and 169.8 degrees for the simulation. These distributions have been obtained by applying a cut to the $\chi^2$ probability, excluding "badly reconstructed" showers (approximately 4% of the total) that exhibit low $\chi^2$ probability.

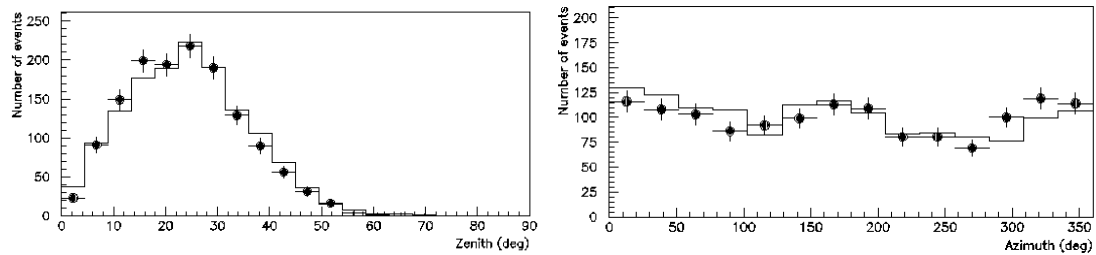

Figure 19: Distribution of the zenith (left) and azimuth (right) angles for EAS detected and reconstructed from all 6 detectors of stations A and B. The experimental data are represented with dots, while the histograms correspond to the simulation estimations..

The limited number of high energy events in the simulation (that trigger both stations A and B) does not allow us to make a detailed MC study of the resolution of the direction reconstruction with 2 HELYCON stations, as it was done when studying the single station performance (e.g. the effect of the primary energy or the distance of the shower core from the stations). Nevertheless, there are enough MC events to present the distribution of the 3d angle difference between the true EAS particle front direction (represented by the direction of the primary particle initiating the shower) inserted in the simulation and the reconstructed EAS direction angle. This distribution is depicted in Figure 20 and has a median value of 2.9 degrees which is better by one degree with respect to the plane shower front approximation (3.9 degrees).

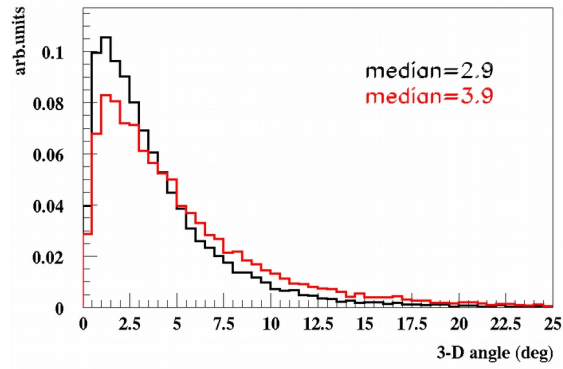

Figure 20: Distribution of the angular difference between the true EAS direction and the EAS particle front as reconstructed by all detectors of stations A and B, taking into account the curvature of the shower front (black) in comparison with the plane shower front approximation (red).

In the case of direction reconstruction of EAS detected from stations A and B simultaneously, the detectors are covering a larger area and the shower front is more extended compared to the local shower front that is detected from a single station, and thus the reconstructed direction approaches the true EAS particle front. In the simulation predictions, this is expressed as a decrease of the rms of the distribution of the difference between true and reconstructed angles for the EAS detected by both stations A and B, in comparison with the same quantity calculated with the triangulation method for each station for exactly the same events. These distributions are depicted in Figure 21, for zenith (left) and azimuth (right) angles, where the direction is estimated by minimizing the $\chi^2$ quantity of Eq. 4 (black) and by employing the triangulation method using station A (red) and station B (green) detectors timing information.

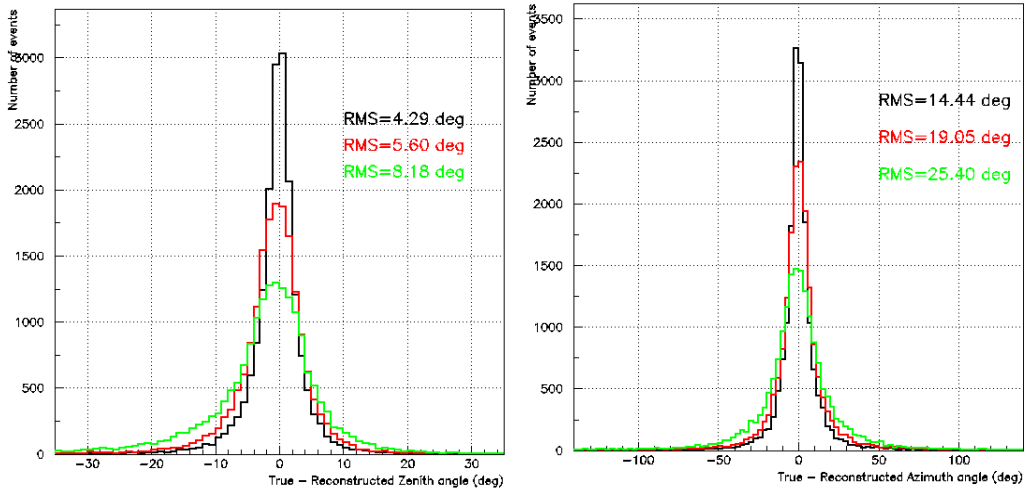

Figure 21: Distribution of the difference between the true and the reconstructed zenith (left) and azimuth (right) angles, as calculated using all 6 detectors timing information (black) and employing the triangulation method using station A (red) and station B (green) timing information.

Figure 22 presents the efficiency of stations A and B for the simultaneous EAS detection in linear (left) and logarithmic (right) scale versus the distance of the EAS impact point from the point that lays at the center of the distance between the stations centers and the EAS primary particle energy. A maximum efficiency of more than

50% is achieved for the highest energy EAS, incident close to the center of the area between the stations. Additionally, the adequacy of the size of the area where the simulated EAS are generated, is proven again, as in the study of the efficiency of the single station (Figure 13).

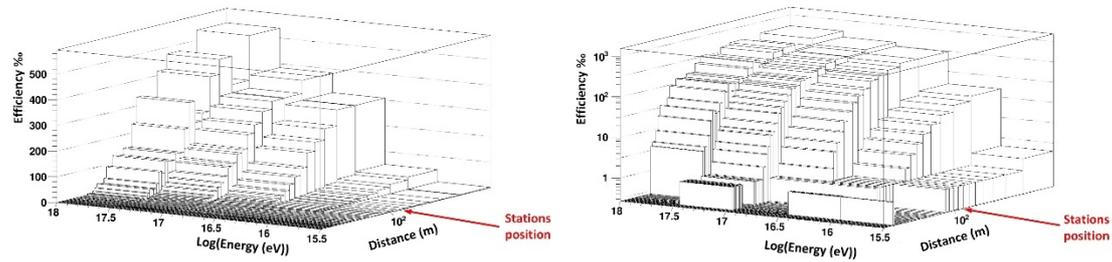

Figure 22: Efficiency of stations A and B in synchronous EAS detection as a function of the EAS impact point distance from the point that lays at the center of the distance between the stations centers and the EAS primary particle energy, in linear scale (left) and logarithmic scale (right).

The shower events triggering more than one station include a substantial fraction of high energy showers that give radio signals, which can be captured by the RF antennas. Although such antennas are supposed to be operated in a radio-quite environment, the employment of a trigger signal from the scintillator detectors stations and the application of quality criteria are sufficient for the detection of the RF signature of EAS in the noisy city environment [5].

## 5 Conclusions

We have presented results for the evaluation of the performance of three autonomous stations concerning the angular resolution and the efficiency in detecting cosmic ray air showers. Event rates, Time over Threshold and collected charge distributions as well as angular distributions from the experimental data were found to be in very good agreement with the predictions of the simulation package. The resolution of the EAS primary particle direction was found to be 3.1º, 5.3º and 3.5º for station A, B and C respectively, while the energy threshold for detecting EAS from each autonomous station is about 10 TeV. Higher energy showers can be selected by combining data from more than one station. Double station coincidence events from stations A and B were analyzed taking into account the curvature of the EAS shower front and the angular resolution of this pair was found to be 2.9º.

## Acknowledgments

This research has been co-financed by the European Union (European Social Fund – ESF) and Greek national funds through the Operational Program "Education and Lifelong Learning" of the National Strategic Reference Framework (NSRF) – Research Funding Program: "THALIS – Hellenic Open University – Development and Applications of Novel Instrumentation and Experimental Methods in Astroparticle Physics" and by the Hellenic Open University Grant No. ΦΚ 228.